\documentstyle{elsart}
\newtheorem{theorem}{Theorem}
\newtheorem{proposition}{Proposition}
\newtheorem{corollary}{Corollary}
\newtheorem{lemma}{Lemma}

\begin{document}

\begin{frontmatter}

\title{Ergodic theorems for algorithmically random points}



\author{Vladimir V. V'yugin\thanksref{label1}}
\thanks[label1]{This research was partially supported by Russian foundation
for fundamental research:  20-01-00203-a.
}

\address{Institute for Information Transmission Problems,
Russian Academy of Sciences,
Bol'shoi Karetnyi per. 19, Moscow GSP-4, 127994, Russia.
e-mail vyugin@iitp.ru}

\date{}


\begin{abstract}
This paper is a survey of applications of the theory of algorithmic randomness to ergodic theory.
We establish various degrees of constructivity for asymptotic laws of probability theory.
In the framework of the KolmogorovТs approach to the substantiation of the probability theory
and information theory on the base of the theory of algorithms, we formulate probabilistic
laws, i.e. statements which hold almost surely, in ``a pointwise''  form, i.e., for
Martin-L\"of random points. It is shown in this paper that the main statement of ergodic
theory -- Birkhoff's ergodic theorem, is non-constructive in the strong (classical) sense,
but it is constructive in some weaker sense -- in terms of Martin-L\"of randomness.
\end{abstract}

\end{frontmatter}

\maketitle

\section{Introduction}

In the framework of the KolmogorovТs approach to the substantiation of the probability theory
and information theory (see Kolmogorov~\cite{Kol65}--\cite{Kol83a})
on the basis of the theory of algorithms, probabilistic
laws, i.e. statements of the form $P\{\omega:A(\omega)\} = 1$, where $A(\omega)$ is some
asymptotic formula (of a probabilistic law),
are presented in ``a pointwise''  form: ``$\omega$ is random $\Rightarrow A(\omega)$''.

Most proofs of such laws, like the strong law of large numbers or law of the iterated logarithm,
stand up to constructive analysis and can be directly translated into the algorithmic form.
An exception is the Birkhoff's ergodic theorem (see Bilingsly~\cite{Bil56}, Krengel~\cite{Kre85}).

In Section~\ref{ergod-1} we analyse the main statement of ergodic theory -- Birkhoff's ergodic theorem, in terms
of algorithmic information theory and Martin-L\"of randomness (see Li and Vitanyi~\cite{LiV97}).
Here the key point is the presence or absence of computable estimates for 
the rate of convergence almost surely for time averages.

In Section~\ref{ergodic-case-1} we show that the classical ergodic theorem for ergodic transformations stands up to 
constructive analysis, since there is a computable estimate for the rate of convergence almost surely 
of time averages in the maximum ergodic theorem. 
The classical proof of this theorem for ergodic measure preserving 
transformations is directly translated into the algorithmic form.

This is not the case in the general position:
we prove that the Birkhoff's ergodic theorem is indeed in some strong sense ``nonconstructive''.

In Section~\ref{non-eff-erg-1} we show that in the case of a not necessarily ergodic measure 
preserving transformation there is no computable estimate for the rate of convergence of 
time averages. Nevertheless, in Section~\ref{erg-prof-1} we show that in the general case, 
for arbitrary measure preserving transformation (not necessary ergodic) a little-known 
Bishop's~\cite{Bis67} proof of the ergodic theorem can be used to obtain the algorithmic 
version of this theorem: time-averaged values of any computable function defined on 
the prefixes of the trajectory of an arbitrary Martin-L\"of random point
converges but there is no computable estimate for the rate of this convergence.

\section{Preliminaries}\label{random-ppp1}

Let $\{0,1\}^*=\cup_n\{0,1\}^n$ be the set of all finite binary sequences,
and $\Omega=\{0,1\}^\infty$ be the set of all infinite binary sequences.
In what follows by a sequence (finite or infinite) we mean the binary sequence, i.e.,
the sequence $\omega_1\omega_2\dots$, where $\omega_i\in\{0,1\}$ for $i=1,2,\dots$.
For any finite or infinite $\omega=\omega_1\dots\omega_n\dots$, we
denote its prefix (initial fragment) of length $n$ as $\omega^n=\omega_1\dots\omega_n$.
We write $x\subseteq y$ if a sequence $y$ is an extension
of a sequence $x$, $l(x)$ is the length of $x$, $\lambda$ is the empty sequence.

Let $\cal R$ be the set of all real numbers,
${\cal R}_+$ be the set of all
nonnegative real numbers, $\cal N$ and $\cal Q$ -- be the sets
of all positive integer numbers and of all rational numbers.

For the basics of computability and algorithmic randomness theory,
see for instance, Rogers~\cite{Rog67}
and Li and Vitanyi~\cite{LiV97}.

We fix the model of computation. Algorithms may be regarded as
Turing machines and so the notion of a program and time of computation
will be well-defined.
Any Turing machine using a program can calculate the values of
a possibly partially defined function $f$ of the type $f:\{0,1\}^*\to \{0,1\}^*$.
Such a function $f$ is called computable or, according to historical tradition,
partial recursive. This means that Turing machine when fed with an input --
a finite sequence $x\in \{0,1\}^*$, transforms it according instructions of some program
to another finite sequence $y$, stops and outputs the result $y=f(x)$
or never stops and outputs no result. In the last case, we say that the result of
computation on the input $x$ is undefined or that the value of $f(x)$ is undefined.

Any algorithm transforms finite objects into finite objects.
Integer and rational numbers (but no reals) are examples of finite objects.
Finite sequences of finite objects are again finite objects.
The main property of finite objects that we use is that
they can be enumerated with positive integers, and therefore, they can be
arguments and values of computable (partial recursive) functions and algorithms.

A function $f$ is computable (or partial recursive) if there is an algorithm
(Turing machine) computing values of $f$. For any input $x$,
the corresponding Turing machine when fed with $x$ stops after several steps
and outputs the result $f(x)$ if $f(x)$ is defined and never stops otherwise.
We call a function $f$ total if $f(x)$ is defined for every $x$.

A set of finite objects is called recursively enumerable if it is the domain
of some computable function. It can be proved that a nonempty set $A$ is
recursively enumerable if and only if it is the range of some total recursive
(computable) function.

Let $A$ be a set of all finite objects of certain type. A function
$f\colon A\rightarrow{\cal R}\cup\{+\infty\}$ is called lower semicomputable if
there is a sequence of total functions $h_n:\{0,1\}^*\to {\cal Q}$ such that
(i) $h_n(x)\le h_{n+1}(x)$ for all $n$ and for all $x$, (ii) the function
$h(n,x)=h_n(x)$ is computable, (iii) $f(x)=\lim_{n\to\infty}h_n(x)$ for
each $x$.

This definition is equivalent to the following one.
A function $f\colon A\rightarrow\cal R\cup \{+\infty\}$ is lower
semicomputable if and only if the set
$$
\{(r,x): x\in\{0,1\}^*,\ r\in{\cal Q}, r<f(x)\}
$$
is recursively enumerable. This means that there is an algorithm which when fed with
a rational number $r$ and a finite object $x$ eventually stops if
$r<f(x)$ and never stops otherwise. In other words, the semicomputability
of $f$ means that if $f(x)>r$ this fact will sooner or later be learned,
whereas if $f(x)\leq r$ we may be for ever uncertain.

A function $f\colon A\rightarrow\cal R\cup \{-\infty\}$ is called upper semicomputable
if there is a sequence of total functions $q_n:\{0,1\}^*\to {\cal Q}$ such that
(i) $q_n(x)\ge q_{n+1}(x)$ for all $n$ and for all $x$, (ii) the function
$q(n,x)=q_n(x)$ is computable, (iii) $f(x)=\lim_{n\to\infty}q_n(x)$ for
each $x$.
This definition is equivalent to the following.
A function $f$ is upper semicomputable if and only if the set
$$
\{(r,x): x\in\{0,1\}^*,\ r\in{\cal Q}, r>f(x)\}
$$
is recursively enumerable.

A function $f:\{0,1\}^*\to {\cal R}$ is called computable if
it is lower semicomputable and upper semicomputable. It can be proved that there
exists an algorithm which, given a finite sequence $x$ and a rational number
$\epsilon>0$, computes a rational approximation of the number $f(x)$ with
accuracy $\epsilon$: given $x$ and a rational $\epsilon>0$ this algorithm
finds an $n$ such that $q_n(x)-h_n(x)<\epsilon$ and outputs 
$h_n(x)$ (or $q_n(x)$) as the result.

The topology on $\Omega$ is generated by intervals
$\Gamma_x=\{\omega\in\Omega:x\subset\omega\}$, where $x$
is a finite binary sequence. The Borel subsets of $\Omega$ can be defined
using these intervals and set theoretic operations.

A probability measure $P$ on $\Omega$ can be defined by the values
$P(x)=P(\Gamma_x)$, where $x\in\{0,1\}^*$. Also, (i) $P(\lambda)=1$
and (ii) $P(x)=P(x0)+P(x1)$ for every $x$. This function is further extended
to all Borel subsets of $\Omega$.

A measure $P$ is computable if the function $x\to P(x)$ is computable.
An example of computable probability measure is
the uniform Bernoulli measure $L$, where $L(\Gamma_x)=2^{-l(x)}$
for any finite binary sequence $x$.

An open subset $U$ of $\Omega$ is called effectively open if it
can be represented as a union of a computable sequence of
intervals: $U=\bigcup_{i=1}^\infty\Gamma_{x_i}$, where
$f(i)=x_i$ is a computable function. A sequence of effectively open sets
$U_n$, $n=1,2,\dots$, is called uniformly effectively open if each set
$U_n$ can be represented as $U_n=\bigcup_{i=1}^\infty\Gamma_{x_{n,i}}$,
where $f(n,i)=x_{n,i}$ is a computable function from $n$ and $i$.

Let $P$ be a computable measure. Martin-L\"of test of randomness with respect to $P$
is an uniformly effectively open sequence $U_n$, $n=1,2,\dots$, of effectively
open sets such that $P(U_n)\le 2^{-n}$ for every $n$. It can be added
the requirement $U_{n+1}\subseteq U_n$ for all $n$.\footnote{It is easy to
that any test $\{U_n\}$ can be redefined as $U'_n=\cup_{i>n}U_i$ such
that $U'_n)\le 2^{-n}$ and $U'_{n+1}\subseteq U'_n$ for all $n$.
}

An infinite binary sequence $\omega$ passes the test $U_n$, $n=1,2,\dots$,
if $\omega\not\in\bigcap U_n$. Otherwise, it is rejected by this test.
A sequence (point) $\omega$ is Martin-L\"of random
with respect to a computable measure $P$ if it passes each Martin-L\"of
test of randomness.


\section{Algorithmic ergodic theory}\label{ergod-1}

We confine our attention to the Cantor probability space
$(\Omega,{\cal F},P)$, where $\Omega$ is the set of all infinite binary sequences,
${\cal F}$ is the collection of all Borel subsets of $\Omega$ generated by
intervals $\Gamma_x=\{\omega\in\Omega:x\subset\omega\}$ and $P$ is a computable measure
on $\Omega$.\footnote{
Hoyrup and Rojas~\cite{HoR2009} showed that any computable probability
space is isomorphic to the Cantor space in both the computable and
measure-theoretic senses. Therefore, there is no loss of generality
in restricting to this case.}

Recall some basic notions of ergodic theory.
An arbitrary measurable mapping of a probability space
into itself is called transformation.
A transformation $T:\Omega\to\Omega$ preserves a measure $P$ on $\Omega$ if
$P(T^{-1}(A))=P(A)$ for all measurable subsets $A$ of the space $\Omega$.
A subset $A$ is called invariant with respect to $T$ if $T^{-1}A=A$ up to
a set of measure 0. A transformation $T$ is called ergodic with respect to $P$
if each subset $A$ invariant with respect to $T$ has measure~0~or~1.

An example of a transformation on $\Omega$ is the (left) shift
$$
T(\omega_1\omega_2\omega_3\dots)=\omega_2\omega_3\dots
$$
If the shift preserves a measure $P$ then this measure is called stationary.
A measure $P$ is called ergodic if the shift is ergodic with respect to $P$.

The uniform measure $L$ is an example of computable stationary and ergodic measure.

\subsection{Poincare's recurrence theorem}\label{poincare-1}

Now, we give an example of the algorithmic analysis in terms of Martin-L\"of randomness
of the well-known statement of ergodic theory -- the Poincare recurrence theorem,
which states as follows:

Let $T$ be a measure $P$ preserving transformation and $E$ be a measurable
subset of $\Omega$.
Then for all $n>0$ the set of all $\omega\in E$ such that $T^n\omega\not\in E$
has measure 0. Equivalently, for almost all $\omega\in E$
it will be $T^n\omega\in E$ for some $n>0$, i.e. trajectory of the point $\omega$
will visit $E$ again. Moreover, it happens infinitely many times.\footnote{
This statement is nontrivial when the measure of the set $E$ is positive.}

We will formulate an algorithmically efficient analogue of this statement
for the uniform measure $L$ on the space $\Omega$ of all infinite binary
sequences: $L(\Gamma_x)=2^{-l(x)}$, and for the shift $T$.
A set $E$ is called effectively closed if
it is the complement of some effectively open set.

An algorithmic version of Poincare's recurrence theorem is presented
in the following theorem
\begin{theorem}\label{kuc-1}
Let $T$ be a shift on $\Omega$ and $E$ be an effectively closed set
of positive measure.
Then for any Martin-L\"of random sequence $\omega\in E$ there will be
$T^n\omega\in E$ for some $n>0$.
Moreover, this is true for infinitely many $n$.
\end{theorem}
Theorem~\ref{kuc-1} will be a direct consequence of the following statement,
which is attributed to Kuchera~\cite{Kuc85}. See also,
Bienvenu et al.~\cite{BDMS2010}.\footnote{Bienvenu et al.~\cite{BDMS2010}
showed that in any computable probability space, a point is Martin-L\"of random
if and only if it is satisfied to the statement of the Poincare's recurrence theorem
for each computable ergodic transformations with respect to effectively closed sets.}

\begin{proposition}\label{kuc-2}
Let $U$ be an effectively open set of uniform measure such that $L(U)<1$.
Then for any infinite sequence $\omega$ Martin-L\"of random
with respect to the measure $L$
and for an arbitrary number $N$ there is a number $n\ge N$
such that $T^n\omega\not\in U$.
\end{proposition}
{\it Proof}. Let $N$ be an arbitrary positive integer number.
By $U^*$ denote the set of all $\omega\in\Omega$ such that
$T^n\omega\in U$ for all $n\ge N$.

Let $L(U)<r$ for some rational number $r$ such that $r<1$.
We represent an effectively open set $U$ as a union of a computable
sequence of pairwise disjoint intervals
$$
U=\cup_i\Gamma_{x_i},
$$
where the finite sequences $x_i$ and $x_j$ do not extend each other.
Denote $U_1=U$ and define
\begin{eqnarray*}
U_2=\cup_{i,j}\Gamma_{x_ix_j},
\\
U_3=\cup_{i,j,s}\Gamma_{x_ix_jx_s}
\end{eqnarray*}
etc. Here $x_ix_j$ is the concatenation of the strings $x_i$ and $x_j$.
$x_ix_jx_s$ is understood in a similar way. We have
\begin{eqnarray*}
L(U_2)=\sum\limits_{i,j}L(\Gamma_{x_ix_j})=
\sum\limits_{i,j}2^{-l(x_i)-l(x_j)}=
\nonumber
\\
=\sum\limits_i2^{-l(x_i)}\sum\limits_j2^{-l(x_j)}<r^2
\end{eqnarray*}
etc. Similarly, we have $L(U_n)<r^n$ for all $n$.

Let $\omega\in U^*$. Denote $\omega'=\omega_{N+1}\omega_{N+2}\dots$.
Since $\omega'=T^N\omega\in U$,
$\omega'=x_i\omega''$ for some $i$ and $\omega''\in\Omega$.
Since $\omega''=T^{N+l(x_i)}\omega\in U$, we have
$\omega''=x_j\omega'''$
for some $j$ and $\omega'''\in\Omega$. Now,
$\omega'=x_ix_j\omega'''$, and then $\omega'\in U_2$.
Similarly, we obtain $\omega'\in U_3$ and etc.

The sequence of the effectively open sets
$\{U_m:m=1,2,\dots\}$
defines a Martin-L\"of test of randomness. It was proved
that $\omega'\in\cap_m U_m$, i.e. $\omega'$ is not Martin-L\"of random.
It is easy to see that in this case the original sequence $\omega$ is also
not random. In particular, $U^*\subseteq\cap_m U_m$. The proposition is proved. 
$\Box$
\bigskip

To prove Theorem~\ref{kuc-1}, one should take $U=\Omega\setminus E$ in Proposition~\ref{kuc-2}.
Since $L(E)>0$, the inequality $L(U)<1$ holds.
By Proposition~\ref{kuc-2} $T^n\omega\not\in U$ (equivalently, $T^n\omega\in E$) for infinitely many $n$.
$\Box$

\subsection{Algorithmically effective convergence}

First, define the notion of computable function of type
$f:\Omega\to {\cal R}$.

A function $h:\Omega\to {\cal R}^+$ is called simple if its
domain $\Omega$ can be represented as a union of finite set of intervals:
$\Omega=\cup_{i=1}^k\Gamma_{x_i}$, where $x_i\in\Xi$, and it takes constant
rational values on each such interval: $h(\omega)=r_i\in {\cal Q}$ for each
$\omega\in\Gamma_{x_i}$ for $1\le i\le k$.
Any simple function is a finite (constructive) object, and so, we can effectively identify
all simple functions and positive integer numbers.\footnote{
This means that there is an algorithm which given any number $n$ can reconstruct
the sets $\{x_1,\dots ,x_k\}$ and $\{r_1,\dots ,r_k\}$. Also, such correspondence
is one-to-one.
}

A function $f:\Omega\to {\cal R}\cup\{+\infty\}$ is called lower semicomputable
if there exists a recursively enumerable non-decreasing sequence $h_n$ of simple
functions: $h_n(\omega)\le h_{n+1}(\omega)$ for all $n$, such that
$f(\omega)=\lim_{n\to\infty}h_n(\omega)$ for each $\omega\in\Omega$.
Similarly, a function $f:\Omega\to {\cal R}$ is called upper semicomputable
if there exists an recursively enumerable non-increasing sequence $h_n$ of simple
functions: $h_n(\omega)\ge h_{n+1}(\omega)$ for all $n$, such that
$f(\omega)=\lim_{n\to\infty}h_n(\omega)$ for each $\omega\in\Omega$.

A function $f:\Omega\to {\cal R}$ is called computable if it is lower semicomptable
and upper semicomptable. It is easy to see that in this case there is an algorithm
which given an infinite sequence $\omega$ and a rational $\epsilon$ computes
a rational approximation of $f(\omega)$ up to $\epsilon$ using some prefix
of $\omega$.

A sequence of functions $f_i:\Omega\to {\cal R}\cup\{+\infty\}$, $i=1,2,\dots$, is called
uniformly lower semicomputable if there exists a recursively enumerable non-decreasing
by $n$ sequence $h_{i,n}$ of simple
functions: $h_{i,n}(\omega)\le h_{i,n+1}(\omega)$ for all $i$ and $n$ such that
for any $i$, $f_i(\omega)=\lim_{n\to\infty}h_{i,n}(\omega)$ for each $\omega\in\Omega$.
The definition of uniformly upper semicomputable sequence of function 
$f_i:\Omega\to {\cal R}$ is similar.
                       
A sequence of computable functions $f_n:\Omega\to{\cal R}$, $n=1,2,\dots$, is called 
uniformly computable it is uniformly lower semicomputable and uniformly 
upper semicomputable.

The algorithmic effective version of convergence in probability and almost surely
of functions $f_n$ of type $\Omega\to{\cal R}$ was considered
by V'yugin~\cite{Vyu97}.

Let $P$ be a probability measure.
A sequence of functions $f_n$ converges to a function $f$ in
probability if for each $\delta>0$
$$
P\{\omega\in\Omega:|f_n(\omega)-f(\omega)|>\delta\}\to 0
$$
as $n\to\infty$. This is equivalent to the fact that there is a function
$m(\delta,\epsilon)$ such that
\begin{eqnarray}
P\{\omega:|f_n(\omega)-f(\omega)|>\delta\}<\epsilon
\label{eff-conv-1}
\end{eqnarray}
for all $n\ge m(\delta,\epsilon)$ for each positive
numbers $\delta$ and $\epsilon$.
The function $m(\delta,\epsilon)$ is called regulator of convergence.

A sequence of functions $f_n$ effectively converges in probability to a function $f$
if there exists a computable regulator of this convergence.

A sequence of functions $f_n$ converges to a function $f$ almost surely
if $\lim_{n\to\infty}f_n(\omega)=f(\omega)$ for almost every $\omega\in\Omega$ (see~\cite{Shi80}).

This definition is equivalent to the following.
A sequence of functions $f_n$ converges to a function $f$ almost surely
if a function $m(\delta,\epsilon)$ (regulator of convergence) exists such that
\begin{eqnarray}
P\{\omega:\sup\limits_{n\ge m(\delta,\epsilon)}
|f_n(\omega)-f(\omega)|>\delta\}<\epsilon
\label{eff-conv-2}
\end{eqnarray}
for every positive rational numbers $\delta$ and $\epsilon$.

A sequence of functions $f_n$ efectively converges a function $f$ almost surely
if there exists a computable regulator of this convergence.

The following simple proposition was proved by V'yugin~\cite{Vyu97}.
\begin{proposition}\label{conf-1}
Let $P$ be a computable measure and a uniformly computable sequence of functions
$f_n:\Omega\to {\cal R}$ effectively converges almost surely
to some function $f$. Then
$$
\lim\limits_{n\to\infty}f_n(\omega)=f(\omega)
$$
for each sequence $\omega$ Martin-L\"of random with respect to $P$.
\end{proposition}
{\it Proof}. By (\ref{eff-conv-2}) we have
$
P\{\omega:\sup\limits_{n,n'\ge m(\delta/2,\epsilon)}
|f_n(\omega)-f_{n'}(\omega)|>\delta\}<\epsilon
$
for every positive rational numbers $\delta$ and $\epsilon$, where $m(\delta,\epsilon)$
is a computable function.

Denote $W_{n,n',j}=\{\omega:|f_n(\omega)-f_{n'}(\omega)|>\frac{1}{j}\}$.
Since the sequence $f_n$ is uniformly computable, this set is effectively open.
Define
$$
V_j=\bigcup_{n,n'\ge m(\frac{1}{2j},2^{-j})}W_{n,n',j}
$$
for all $j$. Define also, $U_i=\bigcup_{j>i} V_j$. Then $P(U_i)\le 2^{-i}$ for all $i$.
Therefore, $\{U_i\}$ is the Martin-L\"of test of randomness.

Assume that $\lim\limits_{n\to\infty}f_n(\omega)$ does not exist for some
$\omega$. Then a number $i$ exists such that $|f_n(\omega)-f_{n'}(\omega)|>1/i$
for infinitely many $n$ and $n'$. For any $j>i$ the numbers
$n,n'\ge m(\frac{1}{2j},2^{-j})$ exist such that $\omega\in W_{n,n',j}\subseteq V_j\subseteq U_i$.
Since $i$ is arbitrary, $\omega\in U_i$ for every $i$.
Hence, the sequence $\omega$ is rejected by the test $\{U_i\}$.
$\Box$
\bigskip


Now we show details of how
Proposition~\ref{conf-1} can be applied to prove that
the strong law of large numbers holds for any sequence
Martin-L\"of random with respect to the uniform measure $L$.

Hoeffding's~\cite{Hoe63} inequality
\begin{eqnarray*}
L\left\{\omega\in\Omega: \left|\frac{1}{n}\sum\limits_{i=1}^n
\omega_i-\frac{1}{2}\right|\ge\delta\right\}\le 2e^{-2n\delta^2}
\end{eqnarray*}
implies
\begin{eqnarray*}
L\left\{\omega\in\Omega: \sup_{n\ge m(\epsilon,\delta)}\left|\frac{1}{n}\sum\limits_{i=1}^n
\omega_i-\frac{1}{2}\right|\ge\delta\right\}\le 2e^{-2n\delta^2+\ln\frac{1}{\delta}}.
\end{eqnarray*}
Define $m(\epsilon,\delta)=\left\lfloor\frac{\ln\frac{2}{\epsilon\delta}}{2\delta^2}\right\rfloor$ for any rational
$\epsilon>0$ and $\delta>0$.
Then we obtain the algorithmically efficient almost sure convergence
\begin{eqnarray*}
L\left\{\omega\in\Omega: \sup_{n\ge m(\epsilon,\delta)}\left|\frac{1}{n}\sum\limits_{i=1}^n
\omega_i-\frac{1}{2}\right|\ge\delta\right\}\le\epsilon
\end{eqnarray*}
for all $\epsilon>0$ and $\delta>0$.

By Proposition~\ref{conf-1}
$
\lim\limits_{n\to\infty}\frac{1}{n}\sum\limits_{i=1}^n\omega_i=\frac{1}{2}
$
holds for each Martin-L\"of random sequence $\omega=\omega_1\omega_2\dots$.

\subsection{Ergodic theorem for Martin-L\"of random sequences for ergodic 
transformations}\label{ergodic-case-1}

Let us define the notion of a computable transformation of
binary sequences. A computable representation of a transformation is
a set $\hat T\subseteq \{0,1\}^*\times \{0,1\}^*$ such that
\begin{itemize}
\item{(i)} the set $\hat T$ is recursively enumerable;
\item{(ii)} for any $(x,y), (x',y')\in\hat T$, if $x\subseteq x'$, then
$y\subseteq y'$ or $y'\subseteq y$;
\item{(iii)}
if $(x,y)\in\hat T$, then $(x,y')\in\hat T$ for all $y'\subseteq y$;
\item{(iv)}
$(x,\lambda)\in\hat T$ for every $x$.
\end{itemize}

A transformation $T$ of the set $\Omega$ is computable if a computable
representation $\hat T$ exists such that (i)-(iv) hold and
$$
T(\omega)=\sup\{y:x\subseteq\omega\&(x,y)\in\hat T)\}
$$
for all infinite sequence $\omega\in\Omega$, where $\sup$
is with respect to the partial ordering $x\subseteq x'$.

Denote $T^0\omega=\omega$, $T^{i+1}\omega=T(T^i\omega)$, so, any
point $\omega\in\Omega$ generates the infinite
trajectory $\omega, T\omega,T^2\omega,\dots$.

Let $P$ be a computable measure, $T$ be a computable ergodic transformation 
preserving the measure $P$. and $f\in L^1$ be a computable function of type 
$\Omega\to {\cal R}$ (observable). 
By $\|f\|$ denote the norm in $L^1$. Assume that
$\sup_\omega |f(\omega)|<\infty$.

Consider the sequence of ergodic time averages $S^f_n$, $n=1,2,\dots$, where
$$
S^f_n(\omega)=\frac{1}{n}\sum\limits_{k=0}^{n-1}f(T^k\omega).
$$

Galatolo et al.~\cite{GHR2010a} and Avigad et al.~\cite{AGT2010} showed that
if the measure preserving transformation $T$ is ergodic then
the time averages $\{S^f_n\}$ effectively converge to
a computable real number $c=\int f(\omega)dP$ almost surely as $n\to\infty$.
We present details of this result for completeness of exposition.
\begin{proposition}\label{erg-aver-1a}
Let $P$ be a computable measure and $T$ be a computable ergodic transformation
preserving the measure $P$. Then the sequence of time averages $\{S^f_n\}$ 
effectively converges $P$-almost surely as $n\to\infty$.
\end{proposition}
{\it Proof}.
We assume without loss of generality that $\int f dP=0$.\footnote{
Replace $f$ with $f-\int f(\omega)dP$ otherwise.
}
The sequence $\|S^f_n\|$ converges to 0 by the classical ergodic theorem.


The maximal ergodic theorem (see Bilingsly~\cite{Bil56}) says that
$$
P\{\omega:\sup\limits_n|S^f_n(\omega)|>\delta\}\le
\frac{1}{\delta}\|f\|
$$
for any measure $P$ and for any measure preserving ergodic transformation $T$.

Given $\epsilon,\delta>0$ compute a $p=p(\delta,\epsilon)$ such that
$\|S^f_p\|\le\delta\epsilon/2$. By the maximal ergodic theorem for $g=S^f_p$
we have
$$
P\{\omega:\sup\limits_n |S^g_n(\omega)|>\delta/2\}\le
\frac{2}{\delta}\|S^f_p\|\le\epsilon.
$$
Now we check that $S^g_n(\omega)$ is not too far from $S_n^f(\omega)$. Expanding $S^g_n(\omega)$,
one can check that
\begin{eqnarray*}
S_n^g(\omega)=\frac{1}{n}\sum\limits_{k=0}^{n-1}g(T^k\omega)=
\frac{1}{np}\sum\limits_{k=0}^{p-1}\sum\limits_{s=0}^{n-1}f(T^{k+s}\omega)=
\frac{1}{np}\left(p\sum\limits_{k=0}^{n-1}f(T^k\omega)\right)+
\\
+\frac{1}{np}\left(\sum\limits_{k=1}^{p-1}(p-k)f(T^{k+n-1}\omega)-
\sum\limits_{k=1}^{p-1}(p-k)f(T^{k-1}\omega)\right).
\end{eqnarray*}
This implies that
\begin{eqnarray*}
\sup\limits_\omega |S_n^g(\omega)-S_n^f(\omega)|\le
\\
\frac{2}{np}\sum\limits_{k=1}^{p-1}(p-k)\sup\limits_\omega|f(\omega)|<
\frac{p-1}{n}\sup\limits_\omega|f(\omega)|\le\delta/2
\end{eqnarray*}
for all $n\ge m(\delta,\epsilon)$, where
$$
m(\delta,\epsilon)=2(p(\delta,\epsilon)-1) r/\delta
$$
and $r$ is the rational number such that $r>\sup\limits_\omega|f(\omega)|$.

If $|S_n^f(\omega)|>\delta$ for some $n\ge m(\delta,\epsilon)$ then
$|S_n^g(\omega)|>\delta/2$. Hence,
$$
P\{\omega:\sup\limits_{n\ge m(\delta,\epsilon)}|S_n^f(\omega)|>\delta\}\le\epsilon,
$$
where $m(\delta,\epsilon)$ is the computable function.
The proposition is proved.
$\Box$
\bigskip

\begin{corollary}
Let $P$ be a computable measure and
$T$ be a computable ergodic transformation preserving the measure $P$.
Then for any sequence $\omega$ Martin-L\"of random with respect to $P$ 
the time average $S^f_n(\omega)$ converges to the number $\int f(\alpha)dP$
as $n\to\infty$.
\end{corollary}
{\it Proof.}
This corollary follows directly from Propositions~\ref{conf-1} and~\ref{erg-aver-1a}.
$\Box$

\subsection{Lack of a computable rate of convergence
in the ergodic theorem for stationary non-ergodic measure}\label{non-eff-erg-1}

In this section, we show that in general case there is no computable rate of
convergence of time averages in the ergodic theorem. We give an example of a computable
measure and a measure preserving transformation for which the convergence of averages 
$S_n^f$ in probability and
almost surely in Birkhoff's theorem is not algorithmically efficient.
\begin{theorem}\label{non-eff-birk-1}
There is a computable measure $P$ and a measure preserving transformation for which there is no computable
rate of the convergence in probability of time averages $S_n^f$
as $n\to\infty$, where $f(\omega)=\omega_1$.\footnote{This limit exists $P$-almost surely
by the classical Birkhoff's ergodic theorem.}

\end{theorem}
{\it Proof}. Let $T$ be the shift on $\Omega$, $f(\omega)=\omega_1$ for $\omega\in\Omega$. 
Then $S_n^f(\omega)=S_n(\omega)=\frac{1}{n}\sum_{i=1}^n\omega_i$.

We will construct a computable stationary 
measure $P$ as a mixture of homogeneous stationary Markov measures
$P_i$, $i=1,2,\dots$. Each measure $P_i$ will be computable and will contain
information about the stopping problem of the universal algorithm.

Let $U(i,\delta,\epsilon)$ be a computable function universal for all computable
functions of two arguments $m(\delta,\epsilon)$.
By universality property of $U(i,\delta,\epsilon)$ for any computable function
$m(\delta,\epsilon)$ a number $i$ exists such that
$m(\delta,\epsilon)=U(i,\delta,\epsilon)$ for all $\delta$ and $\epsilon$(see Rogers~\cite{Rog67}).

We will construct an example of such a measure $P$ using the diagonal argument. Let 
\[
U^s(i,\delta,\epsilon)=
  \left\{
    \begin{array}{l}
      U(i,\delta,\epsilon) \mbox{ if the process of computation terminaes in}\\
\le s \mbox{ steps},
    \\
      \mbox{ undefined, otherwise}.
    \end{array}
  \right.
\]
For any $i$. Define the real number $\alpha_i$ by setting the bits of its binary expansion:
$$
\alpha_i=0.\alpha_{i1}\alpha_{i2}\dots,
$$
where
\[
\alpha_{i,s}=
  \left\{
    \begin{array}{l}
      1 \mbox{, if }u=U^s(i,\frac{1}{4},2^{-(i+1)})\mbox{ is defined end }s>u,
    \\
      0\mbox{ otherwise}.
    \end{array}
  \right.
\]
It is easy to see that the value of each bit of $\alpha_{i,s}$ is
computable by $i$ and $s$. Besides, $\alpha_i>0$ if and only if
the value of $U(i,\frac{1}{4},2^{-(i+1)})$ is defined.
Thus, the real number $\alpha_i$ is an indicator of the stopping
problem at the inputs $\delta=\frac{1}{4}$ and $\epsilon=2^{-(i + 1)}$.

By definition if $\alpha_i>0$ then binary decomposition of the number
$\alpha_i$ consists of a block of zeros followed by ones. In this case
$\alpha_i=2^{-k(i)}$, where $k(i)$ is length of the block of zeros.

Let us define a homogeneous Markov chain for an arbitrary $i$ by specifying
the initial probabilities:
$$
P_i\{\omega_1=0\}=P_i\{\omega_1=1\}=\frac{1}{2}
$$
and the transition probabilities:
$$
P_i\{\omega_{s+1}=0|\omega_s=1\}=P_i\{\omega_{s+1}=1|\omega_s=0\}=\alpha_i
$$
for each $s=1,2,\dots$. It is easy to show that the probability measure $P_i$
generated by the given initial and transition probabilities is stationary.
Moreover, it is computable.

According to the theory of Markov chains (see Shiryaev~\cite{Shi80}), for $\alpha_i>0$
this measure is also ergodic. If $\alpha_i=0$, the measure $P_i$ is concentrated
on only two infinite sequences:
$P_i(0^\infty)=P_i(1^\infty)=\frac{1}{2}$. The sets $\{0^\infty\}$
and $\{1^\infty\}$ are shift-invariant, then if $\alpha_i=0$ the measure $P_i$
is ergodic.

Each measure $P_i$ is computable, and, moreover, there is an algorithm that
computes the value of $P_i(x)$ uniformly in $i$ and $x$.
Define the measure
$$
P(x)=\sum\limits_{i=1}^\infty 2^{-i}P_i(x).
$$
It is easy to prove that the measure $P$ is computable. Since each measure
$P_i$ is stationary, the measure $P$ is also stationary.
It is clear from the definition that this measure is not ergodic.

By definition $S_n^f(\omega)=S_n(\omega)=\frac{1}{n}\sum_{i=1}^n\omega_i$.
By Birkhoff's ergodic theorem applied for the shift, for $P$-almost all
$\omega$ there exists the limit $\lim\limits_{n\to\infty} S_n(\omega)$ as
$n\to\infty$.

Let $m(\delta,\epsilon)$ be an arbitrary computable function,
which is a candidate for the regulator of convergence in probability
of the time averages for the measure $P$. By universality of the function
$U$, an $i$ exists such that $m(\delta,\epsilon)=U(i,\delta,\epsilon)$
for every $\delta$ and $\epsilon$. In this case $\alpha_i>0$.

By the ergodic theorem for Markov processes, the stationary distribution
for the Markov process generated by the measure $P_i$ for $\alpha_i>0$,
is $\pi_0=\frac{1}{2}$ и $\pi_1=\frac{1}{2}$. By the law of large numbers
\begin{eqnarray}
P_i\{\omega:|S_n(\omega)-1/2|<0.01\}\to 1
\label{mark-law-1}
\end{eqnarray}
as $n\to\infty$.

By definition, the number $k(i)$ is equal to the position number of
the last zero in the binary representation of the number $\alpha_i$,
after which there are ones in this representation. It is easy to see that
$\alpha_i=2^{-k(i)}$.

Let us estimate the probabilities $P_i(0^{k(i)})$ and $P_i(1^{k(i)})$.
By definition
\begin{eqnarray*}
P_i(0^{k(i)})=P_i(1^{k(i)})=\frac{1}{2}(1-\alpha_i)^{k(i)-1}>\frac{2}{5}
\end{eqnarray*}
for all sufficiently large $k(i)$.
\footnote{Without loss of generality, we can assume that all steps $s$,
at which some value of the universal function was first defined,
are greater than some fixed value $s_0$.}
Hence,
\begin{eqnarray*}
P_i\{\omega:S_{k(i)}(\omega)=0\mbox{ or }1\}>\frac{4}{5}.
\end{eqnarray*}
By definition $k(i)>m(\frac{1}{4},2^{-(i+1)})$. From here and from (\ref{mark-law-1})
it follows that there is an $n>m(\frac{1}{4},2^{-(i+1)})$ large enough
such that
\begin{eqnarray*}
P_i\{\omega:|S_{k(i)}(\omega)-S_n(\omega)|>\frac{1}{4}\}>\frac{1}{2}.
\end{eqnarray*}
Then $P$-measure of this set is more than
$2^{-i}\cdot\frac{1}{2}=2^{-(i+1)}=\epsilon$, i.e. the numbers $k(i)$ and $n$
do not satisfy the requirement (\ref{eff-conv-1}) for the convergence rate
regulator.

The resulting contradiction proves the theorem.
$\Box$
\bigskip

Since algorithmically efficient convergence almost surely implies algorithmically
efficient convergence in probability, we obtain the following corollary from
the theorem~\ref{non-eff-birk-1}.
\begin{corollary}\label{non-eff-birk-1a}
There is a computable stationary measure $P$ for which there is no computable
regulator for the convergence almost surely for the time averages $S_n^f$.
\end{corollary}

\section{Birghoff's ergodic theorem for Martin-L\"of random sequences}\label{eff-erg-1}

Let $P$ be a measure, $T$ be a measure preserving transformation, and
$f$ be an integrable function (observable).
The classical Birkhoff's ergodic theorem says that for $P$-almost every
points $\omega$
$$
\lim\limits_{n\to\infty}\frac{1}{n}
\sum\limits_{i=0}^{n-1}f(T^i\omega)=\hat f(\omega),
$$
where $\tilde f$ is an integrated and invariant with respect to $T$ function
such that $\int\tilde f(\omega)dP=\int f(\omega)dP$.\footnote{Also, $\hat f(\omega)=E_P[f]$
for almost every $\omega$ if the transformation $T$ is ergodic.
}

Using Bishop's~\cite{Bis67} analysis, V'yugin~\cite{Vyu97},~\cite{Vyu98}
presented an algorithmic version of Birkhoff's pointwise ergodic theorem.
Later this result was extended to a more general spaces by
Hoyrup and Rojas~\cite{HoR2009}, Galatolo et al.~\cite{GHR2010a}, Gacs et al.~\cite{GHR2011}

We will present the proof of this statement for Martin-L\"of random points in
Section~\ref{erg-prof-1}. This proof is based on the Bishop's analysis of
the Birghoff's theorem and on a notion of integral test of randomness,
which will be defined in Section~\ref{int-test-1}.

\subsection{Integral tests of randomness}\label{int-test-1}

For further presentation, we need one more type of a test of randomness -- integral tests.

Let $P$ be a computable measure on $\Omega$. A lower semicomputable
function $f:\Omega\to {\cal R}_+\cup\{+\infty\}$ is called
integral test of randomness with respect to a computable measure
$P$ (integral $P$-test) if
$$
E_P[f]=\int f(\omega)dP\le 1.
$$
Here $E_P$ denotes the mathematical expectation with respect to $P$.

From the definition, for any integral test, the Markov inequality holds:
$$
P\{\omega:f(\omega)>r\}<\frac{1}{r}
$$
for each $r$. In particular, $f(\omega)<\infty$ for almost every $\omega$.

Integral tests can be used to give an equivalent definition of
the Martin-L\"of random sequence.
\begin{theorem}\label{int-1b}
Let $P$ be a computable measure. An infinite binary sequence $\omega $
is Martin-L\"of random with respect to the measure $P$ if and only if
$p(\omega)<\infty$ for each martingal test of randomness $p(\omega)$.
\end{theorem}
{\it Proof}.
Given a martingal test of randomness $p(\omega)$ define
the Martin-L\"of test of randomness
$$
U_m=\{\omega:p(\omega)>2^m\}
$$
for each $m$.

By Markov inequality $P(U_m)\le 2^{-m}$ for each $m$.
Since the lower semicomputable function $p(\omega)$ can be represented
as $p(\omega)=\lim_{n\to\infty}f_n(\omega)$, where $f_n$ is a recursively enumerable
non-decreasing sequence of simple functions, it holds
$$
U_m=\{\omega\in\Omega:\exists n(f_n(\omega)>2^m)\}.
$$
Hence, the set $U_m$ is (uniformly) effectively open.

If $p(\omega)=\infty$ then $\omega\in\cap_n U_n$, i.e., the sequence
$\omega$ is not Martin-L\"of random.

Let us prove the converse statement. Let $\{U_m\}$ be a Martin-L\"of
test of randomness. Define a sequence of characteristic functions
\[
p_m(\omega)=
  \left\{
    \begin{array}{l}
      1 \mbox{ if } \omega\in U_m,
    \\
      0 \mbox{ otherwise}.
    \end{array}
  \right.
\]

The sequence of functions $p_m(\omega)$ is uniformly lower semicomputable,
since $\{U_m\}$ is uniformly effectively open. Define
$$
p(\omega)=\sum\limits_{m=1}^\infty p_m(\omega)
$$
for any $\omega\in\Omega$.
The function $p(\omega)$ is lower semicomputable and
$$
\int p(\omega)dP=\sum\limits_{m=1}^\infty\int p_m(\omega)dP=
\sum\limits_{m=1}^\infty P(U_m)\le
\sum\limits_{m=1}^\infty 2^{-m}=1.
$$
Therefore, the function $p(\omega)$ is the integral $P$-test of randomness.

If $\omega\in\cap_m U_m$ then $p(\omega)=\infty$.
Theorem is proved.
$\Box$

\subsection{Effective version of the Birghoff's ergodic theorem}\label{erg-prof-1}

The effective version of Birkhoff's ergodic theorem will be considered for computable
measure, measure preserving computable transformation, and for computable observable.

The formulation of Birkhoff's ergodic theorem for Martin-L\"of random sequences is obtained
from the original formulation by replacing the expression
``for $P$ -almost every $\omega$'' by ``for each sequence $\omega$ Martin-L\"of random with respect
to the measure $P$''.
\begin{theorem}\label{Bir-1d}
Let $P$ be a computable measure on $\Omega$ and $f$ be an arbitrary computable
integrable function of the type $\Omega\to\cal R_+$.
Then for any measure preserving computable transformation $T$ 
\begin{eqnarray}
\lim\limits_{n\to\infty}\frac{1}{n}\sum\limits_{k=0}^{n-1}f(T^k\omega)=
\tilde f(\omega)
\label{birkhof-1}
\end{eqnarray}
for each Martin-L\"of random sequence $\omega$,
where $\tilde f$ is an integrable invariant with respect to $T$ function
such that $\int\tilde f(\omega)dP=\int f(\omega)dP$.

Moreover, if the transformation $T$ is ergodic then
$\tilde f(\omega)=\int f(\alpha)dP$ for each random $\omega$.
\end{theorem}
{\it Proof}.
For an arbitrary infinite sequence $\omega$, consider the time average
$$
s_m(\omega)=S_{m+1}^f(\omega)=\frac{1}{m+1}\sum\limits_{k=0}^m f(T^k\omega).
$$
For convenience, in the further reasoning, we assume that $s_{-1}(\omega)=0$.

Assume that $\int |f(\omega)|dP\le M$, where $M$ is a positive integer number.

If the limit $\lim\limits_{m\to\infty}s_m(\omega)$ does not exist then
there exist the rational numbers $\alpha<\beta$ such that $-M<\alpha<\beta<M$ and
\begin{eqnarray*}
\liminf\limits_{m\to\infty} s_m(\omega)<\alpha<\beta<
\limsup\limits_{m\to\infty} s_m(\omega).
\end{eqnarray*}
The converse is also true.

Let $\alpha$ and $\beta$ be rational numbers such that
$-M<\alpha<\beta<M$. Define the boundary crossing function
$\sigma_n(\omega|\alpha,\beta)$ as follows.
The value $\sigma_n(\omega|\alpha,\beta)$ is equal to the number of
upward intersections of the interval $(\alpha,\beta)$ by the sequence
$s_0(\omega),s_1(\omega),~\dots,~s_n(\omega)$.
More precisely, we define
\begin{eqnarray*}
u_0=0,
\\
u_1=\min\{m:m\ge u_0,s_m(\omega)<\alpha\},
\\
v_1=\min\{m:m>u_1,s_m(\omega)>\beta\},
\\
\dots~~~~~~~~~~~~~~~~~~~
\\
u_i=\min\{m:m>v_{i-1},s_m(\omega)<\alpha\},
\\
v_i=\min\{m:m>u_i,s_m(\omega)>\beta\},
\\
\dots~~~~~~~~~~~~~~~~~~~
\\
u_k=\min\{m:m>v_{k-1},s_m(\omega)<\alpha\},
\\
v_k=\min\{m:m>u_k,s_m(\omega)>\beta\}.
\end{eqnarray*}
Define the function
\[
\sigma_n(\omega|\alpha,\beta)=
  \left\{
    \begin{array}{l}
      0 \mbox{ if } v_1>n,
    \\
      \max\{k:v_k\le n\} \mbox{ if } v_1\le n.
    \end{array}
  \right.
\]
The value $\sigma_n(\omega|\alpha,\beta)$ is equal to the maximum
number of upward crossings of the interval
$(\alpha,\beta)$ by the average $s_m(\omega)$ for $m=0,1,\dots n$.
The function $\sigma_n(\omega|\alpha,\beta)$ is uniformly lower semicomputable
with respect to the arguments $n$, $\alpha$, $\beta$.

It is easy to see that the limit $\lim\limits_{m\to\infty} s_m(\omega)$
does not exist if and only if $\sup_n\sigma_n(\omega|\alpha,\beta)=\infty$
for some $\alpha$ and $\beta$ such that $\alpha<\beta$.

We temporarily fix the infinite sequence $\omega$, as well as the positive integer
number $n$ and the rational numbers $\alpha$ and $\beta$ such that $\alpha<\beta$.

Consider the non-relative deviations
\begin{eqnarray*}
a(u,\omega)=\sum\limits_{s=0}^u (f(T^s\omega)-\alpha),
\\
b(v,\omega)=\sum\limits_{s=0}^v (f(T^s\omega)-\beta).
\end{eqnarray*}
It will be convenient for us to assume that $a(-1,\omega)=0$.


Oscillation of relative frequencies entails oscillation of non-relative
deviations.

A sequence $d=\{u_1,v_1,~\dots,~u_k,v_k\}$ of integer numbers
is called admissible if
$$
-1\le u_1<v_1\le u_2<v_2\le\dots\le u_k<v_k\le n.
$$
The number of pairs in the admissible sequence $d$ will be denoted by $m_d$
($m_d = k$) and will be called its length.

For any admissible sequence
$$
d=\{s_1,t_1,\dots ,s_k,t_k\},
$$
consider the cumulative sum of the differences of non-relative deviations:
\begin{eqnarray*}
S(d,\omega)=\sum\limits_{j=1}^k (b(t_j,\omega)-a(s_j,\omega)).
\end{eqnarray*}
The key role in the proof of the theorem is played by the following combinatorial
lemma on lengthening an admissible sequence without decreasing the cumulative sum.

\begin{lemma}\label{komb-1}
For each admissible sequence $q$, there exists an admissible sequence $d$
such that $m_d\ge\sigma_n(\omega|\alpha,\beta)$ and $S(d,\omega)\ge S(q,\omega)$.
\end{lemma}
{\it Proof}. Denote by $N=\sigma_n(\omega|\alpha,\beta)$
the maximum number of intersections of the interval $(\alpha,\beta)$ by
a sequence of averages $s_0(\omega),\dots~,s_n(\omega)$. Let
$$
p=\{-1<u_1<v_1<u_2<v_2<\dots <u_N<v_N\le n\}
$$
be that admissible sequence of length $N$ by which the value
$\sigma_n(\omega|\alpha,\beta)$
was determined: $N=\sigma_n(\omega|\alpha,\beta)$.

It suffices to prove that for any admissible sequence
$q$ of length $m_q<N$ an admissible sequence $d$ exists such that
$m_d=m_q+1$ and $S(d,\omega)\ge S(q,\omega)$. We will use
some elements of the sequence $p$ to construct such $d$.

Let an admissible sequence $q$ be given:
$$
-1\le s_1<t_1\le s_2<t_2\le\dots\le s_m<t_m\le n,
$$
where  $m=m_q<N$. We expand it by one pair of elements.
Consider an auxiliary element $s_{m+1}=n$.
Since $m+1\le N$, the element $v_{m+1}$ is presented in
the sequence $p$.

Besides, $v_{m+1}\le n=s_{m+1}$. Therefore, there is the smallest $i$
such that $v_i\le s_i$. If $i=1$ then define
\begin{eqnarray}
d=\{u_1,v_1,s_1,t_1,~\dots,~s_m,t_m\}.
\label{seq-1}
\end{eqnarray}
The length of the admissible sequence $q$ has increased by one.

Consider the case $i>1$. Then $v_{i-1}>s_{i-1}$ and the inequality
\begin{eqnarray*}
s_{i-1}<v_{i-1}<u_i<v_i\le s_i
\end{eqnarray*}
is valid. If $u_i<t_{i-1}$ define
\begin{eqnarray}
d=\{s_1,t_1,~\dots,~s_{i-1},v_{i-1},u_i,t_{i-1},~\dots,~s_m,t_m\}.
\label{seq-2}
\end{eqnarray}
If $u_i\ge t_{i-1}$ define for $i\le m$
\begin{eqnarray}
d=\{s_1,t_1,\dots~,s_{i-1},t_{i-1},u_i,v_i,s_i,t_i,\dots~,s_m,t_m\},
\label{seq-3}
\end{eqnarray}
for $i=m+1$ define
\begin{eqnarray}
d=\{s_1,t_1,~\dots,~s_m,t_m,s_{m+1},t_{m+1}\}.
\label{seq-4}
\end{eqnarray}
The sequence $d$ is admissible and its length has increased by one:
$m_d=m_q+1$. It remains to check how the cumulative sums have changed
for all variants of this definition.

By definitions (\ref{seq-1}), (\ref{seq-3}) and (\ref{seq-4})
$$
S(\omega,d)=S(\omega,q)+b(v_i,\omega)-a(u_i,\omega)
$$
and the added term is positive.
If $d$ was defined by (\ref{seq-2}) then
$$
S(\omega,d)=S(\omega,q)+b(v_{i-1},\omega)-a(u_i,\omega).
$$
By the definition of the sequence $\{u_1,v_1,~\dots,~u_N,v_N\}$
$s_{v_{i-1}}(\omega)>\beta$ and $s_{u_i}(\omega)<\alpha$.
Then
$b(v_{i-1},\omega)>0>a(u_i,\omega)$ and
the added term is also positive.

Therefore, in both cases the cumulative sum increases:
$
S(\omega,d)>S(\omega,q).
$
Lemma is proved.
$\Box$
\bigskip

Let $d=\{s_1,t_1,~\dots,~s_m,t_m\}$ be an admissible sequence of length
$m_d=m$ and $S(\omega,d)$ be the corresponding cumulative sum.

Let us apply the transformation $T$ to the sequence $\omega $ and show how
the cumulative sum changes.
First, when $s_i\ge 0$, the following changes occur:
\begin{eqnarray*}
a(s_i,\omega)=a(s_i-1,T\omega)+f(\omega)-\alpha,
\nonumber
\\
b(t_i,\omega)=b(t_i-1,T\omega)+f(\omega)-\beta.
\end{eqnarray*}
From this and from the definition of the cumulative sum,
we obtain
\begin{eqnarray}
S(\omega,d)=S(T\omega,d')+a-(\beta-\alpha)m_d,
\label{kum-sum-ch-1}
\end{eqnarray}
where
$$
d'=\{s_1-1,t_1-1,~\dots,~s_m-1,t_m-1\}
$$
if $s_1\ge 0$, and
$$
d'=\{-1,t_1-1,s_2-1,t_2-1,~\dots,~s_m-1,t_m-1\}
$$
if $s_1=-1$ and $t_1>0$. If $s_1=-1$ and $t_1=0$ then
$$
d'=\{s_2-1,t_2-1,~\dots,~s_m-1,t_m-1\}.
$$
In the sum (\ref{kum-sum-ch-1}), $a=0$ if $s_1\ge 0$ and $a=f(\omega)-\alpha$
if $s_1=-1$.

Let us introduce the lower semicomputable function
\begin{eqnarray*}
\lambda_n(\omega)=\sup\{S(\omega,d): d\mbox{ -- admissible sequence}\}.
\end{eqnarray*}
Then by (\ref{kum-sum-ch-1}) we obtain
\begin{eqnarray}
S(\omega,d)\le\lambda_n(T\omega)+(f(\omega)-\alpha)^+-(\beta-\alpha)m_d,
\label{ineq-l-1}
\end{eqnarray}
where $h^+=\max\{h,0\}$.

By Lemma~\ref{komb-1} for any admissible sequence
$q$, an admissible sequence $d$ exists such that
$m_d\ge\sigma_n(\omega |\alpha,\beta)$ and
$S(\omega,q)<S(\omega,d)$. Then by (\ref{ineq-l-1})
\begin{eqnarray}
S(\omega,q)<S(\omega,d)\le
\nonumber
\\
\le\lambda_n(T\omega)+(f(\omega)-\alpha)^+-(\beta-\alpha)
\sigma_n(\omega|\alpha,\beta),
\label{ineq-l-2}
\end{eqnarray}
We take in (\ref{ineq-l-2}) maximum by $q$ and get
\begin{eqnarray*}
\lambda_n(\omega)\le\lambda_n(T\omega)+(f(\omega)-\alpha)^+-(\beta-\alpha)
\sigma_n(\omega|\alpha,\beta).
\label{ineq-l-3}
\end{eqnarray*}
Therefore,
\begin{eqnarray}
(\beta-\alpha)\sigma_n(\omega|\alpha,\beta)\le (f(\omega)-\alpha)^++
\lambda_n(T\omega)-\lambda_n(\omega).
\label{ineq-l-4}
\end{eqnarray}
Integrating the inequality (\ref{ineq-l-4}), we obtain
\begin{eqnarray}
\int (\beta-\alpha)\sigma_n(\omega|\alpha,\beta)dP\le
\int (f(\omega)-\alpha)^+ dP.
\label{ineq-l-5}
\end{eqnarray}
Here, we use the assumption that the transformation $T$
preserves the measure $P$. This assumption implies that
$$
\int\lambda_n(T\omega)dP=\int\lambda_n(\omega)dP.
$$
Since the integral of the function $|f(\omega)|$ is bounded by the
number $M$,
$$
\int (f(\omega)-\alpha)^+dP\le 2M.
$$
Define
$$
\sigma(\omega|\alpha,\beta)=\sup_n\sigma_n(\omega|\alpha,\beta).
$$
It is easy to see that this function is lower semicomputable.
In addition, since
$$
\sigma_n(\omega|\alpha,\beta)\le\sigma_{n+1}(\omega|\alpha,\beta)
$$
for all $n$, this function is integrable and by (\ref{ineq-l-5})
$$
\int (2M)^{-1}(\beta-\alpha)\sigma(\omega|\alpha,\beta)dP\le 1
$$
for each $\alpha$ and $\beta$ such that $\alpha<\beta$.

By averaging the quantity $\sigma(\omega|\alpha,\beta)$ we can
define an integral test of randomness as follows. Let the computable functions
$\alpha(i)$ and $\beta(i)$ enumerate the set of all pairs of rational numbers
$
\{(\alpha,\beta):-M<\alpha<\beta<M\}.
$
Define
$$
p(\omega)=\frac{1}{2M}\sum\limits_{i=1}^\infty
\frac{1}{i(i+1)}(\beta(i)-\alpha(i))\sigma(\omega|\alpha(i),\beta(i)).
$$
By definition the function $p(\omega)$ is lower semicomputable and
$$
\int p(\omega)dP\le 1,
$$
i.e., it is an integral test of randomness with respect to the measure $P$.
In addition, as previously noted, if the limit of averages
$\lim\limits_{n\to\infty}s_n(\omega)$ does not exist then rational numbers
$\alpha$ and $\beta$ exist such that $\alpha<\beta$ and
$\sigma(\omega|\alpha,\beta)=\infty$. In this case $p(\omega)=\infty$.

Therefore, for any infinite binary sequence $\omega$ the implication
is true:
\begin{eqnarray*}
p(\omega)<\infty\Rightarrow\lim\limits_{n\to\infty}
\frac{1}{n}\sum\limits_{k=0}^{n-1}f(T^k\omega)\mbox{ exists}.
\end{eqnarray*}
The main part of the theorem is proved.

Denote by $\tilde f(\omega)$ the limit of averages (\ref{birkhof-1}).
It is easy to see that $\tilde f(\omega)$ is defined and
$\tilde f(T\omega)=\tilde f(\omega)$ for almost every
$\omega$. 

If the transformation $T$ is ergodic then $\tilde f(\omega)=c$
for $P$-almost all $\omega$, where $c=\int f(\omega)dP$ is the constant.

We need to prove the following statement.
\begin{lemma}
$\tilde f(\omega)=\int f(\omega)dP$ for each sequence $\omega$ random
with respect to the measure $P$.
\end{lemma}
{\it Proof}.
Assume that this assertion is violated. Then a random sequence $\omega$
exists such that $\tilde f(\omega)=d\not =c$. Take rational numbers $r_1$ and $r_2$
such that $r_1<d<r_2$ and $c\le r_1$ or $c\ge r_2$ and define
\begin{eqnarray*}
S_n=\{\alpha:r_1<s_n(\alpha)<r_2\},
\\
\bar S_n=\{\alpha:r_1\le s_n(\alpha)\le r_2\}.
\end{eqnarray*}
Since the limit (\ref{birkhof-1}) is equal to $c$ almost surely,
$P(\bar S_n)\to 0$ as $n\to\infty$. The function $P(\bar S_n)$ is 
upper semicomputable (by $n$), since
\begin{eqnarray*}
r>P(\bar S_n)\Leftrightarrow 1-r<P\{\alpha:r_1>s_n(\alpha)
\mbox{ or }r_1<s_n(\alpha)\}.
\end{eqnarray*}
Therefore, using any $m$, we can efficiently find an $n\ge m $
such that $P(\bar S_n)<2^{-m}$.

By definition the set $S_n$ is effectively open. We have
$P(S_n)\le P(\bar S_n)<2^{-m}$. Define $U_m=S_n$ for such $n$.
The family $\{U_m\}$ of effectively open sets defines the Martin-L\"of
test of randomness with respect to the measure $P$.

It holds $\omega\in\cap_{m=1}^{\infty} U_m$, i.e., $\omega$ is not Martin-L\"of random.

The resulting statement proves the lemma and Theorem~\ref{Bir-1d}.
$\Box$
\bigskip

Later a converse result was obtained by Franklin and Towsner~\cite{FrT14}.
Using the cutting and stacking method, they showed that for every infinite sequence
$\omega$, which is not Martin-L\"of random with respect to the uniform measure $L$
the measure preserving transformation $T$ can be constructed such that the limit
(\ref{birkhof-1}) does not exist.




\begin{thebibliography}{99}

\bibitem{AGT2010}
Avigad, J., Gerhardy, P., Towsner H.: Local stability
of ergodic averages.
Transactions of the American Mathematical Society {\bf 362} 1
(2010) 261Ц288

\bibitem{BDMS2010}
Bienvenu, L., Day, A. Mezhirov, I.,  Shen, A.
Ergodic-type characterizations of Martin-L\"of randomness.
In 6th Conference on Computability in Europe (CiE 2010), volume 6158 of Lecture
Notes in Comput. Sci., pages 49--58. Springer, Berlin, 2010

\bibitem{Bil56}
Billingsley, P,: Ergodic Theory and Information. Wiley, 1956.


\bibitem{Bis67} Bishop, E. Foundation of Constructive Analysis.
    New York: McGraw-Hill, 1967


\bibitem{FrT14}
Franklin, J.N.Y., Towsner, H.: Randomness and non-ergodic systems.
Mosc. Math. J., 2014, {\bf 14} 4, 711--744


\bibitem{GHR2011} Gacs, P., Hoyrup, M., Rojas, C.: Randomness
    on Computable Probability Spaces-A Dynamical Point of View.
    Theory of Computing Systems {\bf 48} 3 (2011) 465--485



\bibitem{Hoe63}
Hoeffding, W.: Probability inequalities for sums of bounded random
variables. Journal of the American Statistical Association {\bf 58} 301
(1963) 13-Ц30

\bibitem{HoR2009}
Hoyrup, M., Rojas, C.: Computability of probability
measures and Martin-L\"of randomness over metric spaces. Information and
Computation {\bf 207} 7 (2009) 830-Ц847

\bibitem{GHR2010a}
Galatolo, S., Hoyrup, M., Rojas, C.:
Computing the speed of convergence of ergodic averages and
pseudorandom points in computable dynamical systems
Computability and Complexity in Analysis (CCA 2010)
EPTCS {\bf 24} (2010) 7-Ц18 doi:10.4204/EPTCS.24.6


\bibitem{Kol65} Kolmogorov, A.N. Three approaches to the quantitative definition of information.
\emph{Problems Inform. Transmission}, \textbf{1}(1), 1965, 1--7.

\bibitem{Kol69} Kolmogorov. A.N. On the logical foundations of information theory and probability theory.
\emph{Problems Inform. Transmission}, \textbf{5}(3), 1969, 1--4.

\bibitem{Kol83} Kolmogorov, A.N. Combinatorial foundations of information theory and the calculus of probabilities.
\emph{Russian Math. Surveys}, \textbf{38}(4), 1983, 29--40.

\bibitem{Kol83a} Kolmogorov, A.N. On logical foundations of probability. \emph{Lecture Notes in Mathematics},
\textbf{1021}, 1983, 1--5.

\bibitem{Kre85}
Krengel, U. Ergodic Theorems, Berlin, New York: de Cruyter 1984.

\bibitem{Kuc85}
Kucera A. Measure, $\Pi_1^0$ classes, and complete extensions of PA.
Lecture Notes in Mathematics. {\bf 1141} 1985, 245--259.

\bibitem{LiV97}
Li, M., Vit{\'a}nyi, P.
An Introduction to Kolmogorov Complexity and Its Applications,
Springer-Verlag. New York 1997.

\bibitem{Rog67}
Rogers, H. Theory of Recursive Functions and Effective Computability,
New York: McGraw-Hill 1967.



\bibitem{Shi80}
Shiryaev, A.N. Probability, Berlin: Springer 1980.

\bibitem{Vyu97}
V'yugin, V.V.: Effective Convergence in Probability and an Ergodic
Theorem for Individual Random Sequences. Theory Probab. Appl. {\bf 42} (1) 1998,
39-Ц50.

\bibitem{Vyu98}
V'yugin, V.V.: Ergodic theorems for individual random sequences.
Theoretical Computer Science {\bf 207} (4) 1998, 343--361.




\end{thebibliography}
\end{document}